\documentclass[a4paper,conference]{IEEEtran}
\IEEEoverridecommandlockouts
\usepackage{cite}
\usepackage{amsmath,amssymb,amsfonts}
\usepackage{algorithmic}
\usepackage{graphicx}
\usepackage{textcomp}
\usepackage{xcolor}

\usepackage{todonotes}
\usepackage{float}
\usepackage{enumitem}
\usepackage{scrextend}
\usepackage{slashbox}
\usepackage{pict2e}
\usepackage{caption}
\usepackage{fontawesome}
\usepackage{tabu}
\usepackage{array}

\usepackage{booktabs}
\usepackage{rotating}

\usepackage{tabularx}
\newcolumntype{Y}{>{\centering\arraybackslash}X}
\newcommand*\rot{\rotatebox{90}}

\def\BibTeX{{\rm B\kern-.05em{\sc i\kern-.025em b}\kern-.08em
    T\kern-.1667em\lower.7ex\hbox{E}\kern-.125emX}}
\begin{document}

\title{Network intrusion detection systems for in-vehicle network - Technical report \\
}

\author{\IEEEauthorblockN{Guillaume Dupont\IEEEauthorrefmark{1}, Jerry den Hartog\IEEEauthorrefmark{1}, Sandro Etalle\IEEEauthorrefmark{1} and Alexios Lekidis\IEEEauthorrefmark{2}
\IEEEauthorblockA{Dept. of Mathematics and Computer Science, Eindhoven University of Technology}
\IEEEauthorblockA{Forescout Technologies, Eindhoven}
Email: \IEEEauthorrefmark{1}\{g.f.c.dupont,j.d.hartog,s.etalle\}@tue.nl,
\IEEEauthorrefmark{2}alexis.lekidis@forescout.com}}

\maketitle

\begin{abstract}
Modern vehicles are complex safety critical cyber physical systems, that are connected to the outside world, with all security implications that brings.
To enhance vehicle security several network intrusion detection systems (NIDS) have been proposed for the CAN bus, the predominant type of in-vehicle network. 
The in-vehicle CAN bus, however, is a challenging place to do intrusion detection as messages provide very little information; interpreting them requires specific knowledge about the implementation that is not readily available.
In this technical report we collect how existing solutions address this challenge by providing an organized inventory of various CAN NIDSs present in the literature, categorizing them based on what information they extract from the network and how they build their model.

\end{abstract}


\section{Introduction}

%
%

Vehicles are becoming more intelligent, offering increasing numbers of innovative applications covering different functionalities ranging from vehicle control to telematics and advanced driver assistance systems.  To achieve this they implement more than a 100 million lines of code, running on micro controllers (called electronic control units, ECUs) spread over the entire vehicle and they are connected to the outside world; to personal devices, to VANETs and the Internet.
In short, a modern vehicle is a complex cyber physical network rather than purely a mechanical device.

The increased connectivity and many functionalities, while offering many benefits, also come with evident security risks. The vehicle complexity offers a large attack surface. Researchers have already demonstrated the ability to remotely take over the control of diverse vehicles at speed~\cite{net_checkoway2011comprehensive,net_miller2015remote,net_miller2016, net_nie2017}. Such (Remote) attacks could have life threatening consequences.

Implementing security measures in a safety critical complex cyber physical network is difficult as one has to guarantee the security measures do not impact the existing functionality. Invasive solutions that could impact availability of safety critical functions or require a major redesign of the vehicle or its components will not be acceptable to the manufacturer.
Network intrusion detection systems (NIDS) provide a non-invasive security measure that is well-established in other fields (general IT, ICS). 
But how well do such approaches translate to the automotive setting?
In literature we find a number of NIDS proposals for the predominant type of in-vehicle network: the Controller Area Network (CAN) bus~\cite{net_abbott-mccune2017,net_gmiden2016,net_moore2017,net_japkowicz2015,net_otsuka2014,net_waszecki2017automotive,net_cho2016,net_muter2011entropy,net_marchetti2016,net_stabili2017,net_taylor2016,net_miller2014survey,net_song2016intrusion,net_ujiie2016,net_lee2017,net_marchetti2017,net_kang2016,net_larson2008approach, net_hoppe2011,net_muter2010structured,net_muter2011} that address of the particularities of this environment in different ways.

In this technical report we provide an organized inventory of present CAN NIDs proposals, categorizing them using dimensions suitable for the in-vehicle CAN bus setting.

The remainder of this paper is structured as follows:
after treating the CAN protocol, known attacks on CAN and NIDS in general in Section~\ref{sec:can} we address the threat landscape in Section~\ref{sec:threatmod}. Next we survey existing in-vehicle NIDS in Section~\ref{sec:survey}.

\section{Preliminaries}
In this section we provide the background information relevant to our study.
We first give an overview of the Controller Area Network (CAN) protocol, followed by the CAN bus attacks published, and finally we cover some basic terminology regarding NIDS.

\subsection{Controller Area Network}
\label{sec:can}

CAN is a message-oriented transmission protocol proposed by Robert Bosch in 1986. It was originally designed for the automotive industry, in an attempt to reduce the wiring complexity of automobiles \cite{net_introtocan}. Due to its low cost, simplicity, deterministic resolution of contention and resilience to electromagnetic interference, CAN is today not only the \textit{de facto} standard protocol for in-vehicle communications, but also widely used in hospitals, factories and plant controls \cite{net_understanding_can}.

CAN is a multicast communication protocol, based on a multi-master access scheme \cite{net_canproto_understandingcan}. In the automotive context, the Electronic Control Units (ECU) broadcast their CAN frames onto the bus. There is no addressing scheme with CAN: each frame is assigned a unique identifier, known as the ``arbitration ID'' or ``CAN ID'', which defines both the content and the priority of the frame. With 11 bits in the identifier field, the CAN ID can range from 0x000 to 0x7FF. 

\begin{figure}[h]
 \centering
 \includegraphics[width=\columnwidth]{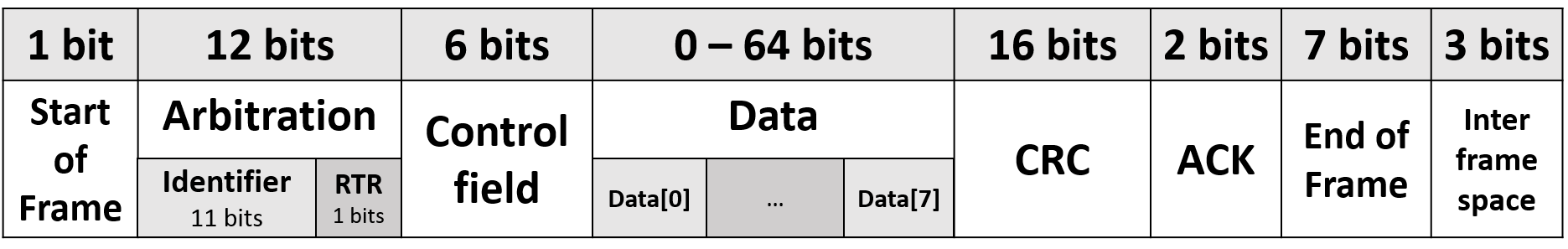}
 \caption{CAN frame structure}
 \label{fig:canframestructure}
\end{figure}

CAN frames are composed of several fields, as depicted in Fig.~\ref{fig:canframestructure}. According to the CAN specifications, four different types of frame are used to control message transfers on the bus \cite{net_can_spec}. \textit{Data frames} (RTR=0) are typically used to carry up to 8 bytes of data sent by an ECU. Moreover a node on the bus can also send a  \textit{remote frame} (RTR=1, 0 bit data field) to request information from another one.

\textit{Overload frames} (specified in the control field) can request an extra delay between two data or remote frames. 

Finally, any node on the bus can invalidate a frame being sent by sending an \textit{error frame}, when it detects a problem with it.

While an ECU may broadcast multiple CAN IDs, each CAN ID is bound to a single ECU; 

two ECUs cannot send data frames with the same CAN ID. Every time a frame is transmitted, all ECUs on the bus will receive it, and will determine, based on the CAN ID whether they should accept and further process the message \cite{net_canoverview}.

In the automotive context, vendors rely on higher level protocols on top of CAN to define the format of the CAN frames' payload. Generally speaking, they use two types of data frames: normal messages and diagnostic messages \cite{net_miller2015remote}. Normal messages follow a proprietary format and are the ones transmitted by the ECUs during regular operations. Diagnostic messages are defined according to a diagnostic communication protocol such as Unified Diagnostic Services (ISO 14229-1): they are special messages normally sent by mechanic's tool (or ``tester'') during maintenance operations. Depending on the services implemented on an ECU, diagnostic messages can be used to perform various actions such as querying information or updating its firmware.

The CAN protocol offers a robust arbitration mechanism that handles the conflict when two or more nodes try to transmit a frame at the same time.
As previously mentioned, the CAN ID of a frame determines its priority: the lower the ID value, the higher its priority. Consequently the ECU sending the frame with the lowest CAN ID value wins the arbitration and has access to the bus. The other ECUs wanting to transmit data wait until the bus is free before trying again. The advantage of the arbitration mechanism is twofold: it guarantees that neither information nor time is lost \cite{net_can_spec} and that eventually all ECUs can have access to the bus \cite{net_cantuto}.

Additionally CAN includes a fault confinement mechanism which removes faulty ECUs from the bus. A node can be in one of these three states: \textit{error active} when functioning properly, \textit{error passive} when suspected of faulty behavior, or \textit{bus off} when considered corrupted \cite{net_understanding_can}. When too many errors are detected, ECUs will change state and ultimately disconnect themselves from the network \cite{net_canoverview}. 

This mechanism protects the health of the bus by preventing faulty ECUs from disturbing communications or impacting network performance~\cite{net_canproto_understandingcan}.

The CAN protocol and its application has a few important security implications. First of all, car makers deploy the reliable CAN protocol on high-speed buses inside their vehicles to interconnect safety-critical ECUs. However, its simplicity implies that  many common security measures such as communication encryption or ECU authentication are not possible. In addition, the broadcasting nature of the protocol allows anyone on the bus to read and send messages, which can have critical consequences as explained in the next section.

\subsection{CAN bus attacks}
\label{sec:attacks}

As demonstrated in previous research \cite{net_checkoway2011comprehensive,net_miller2015remote, net_rouf2010,net_nie2017}, an attacker has a board range of entry points which can be leveraged to gain access to the CAN bus. 
Due to the nature of CAN communications (unencrypted and broadcast communications) an attacker can perform various actions once she gained access to the bus.
The attacker can start with some sort of initial reconnaissance by listening to the messages sent onto the bus and performing a \textit{fuzzing attack} by sending messages with random CAN IDs and payload values and observing the reaction. This attack can allow an attacker to learn about the architecture and the behavior of the vehicle and its ECUs. 

As demonstrated in \cite{net_miller2015remote}, an attacker can also send a diagnostic message in order to open a diagnostic session on an ECU, allowing her to perform specific actions, depending on the services implemented on that ECU.
Diagnostic messages are extremely powerful.
They are, for example, leveraged by Miller and Valasek in \cite{net_miller2013} to perform various actions on a Toyota Prius and a Ford Escape.

The attacker can try to influence an ECU by crafting frames with the CAN ID and payload of his choice. With sufficient knowledge about the payload, the attacker can craft CAN frames in such way that the receiving ECU will behave according to the attacker's wishes. She can for instance make the instrument cluster display an arbitrary speed or steer the wheels in a given direction.


If the attacker aims to prevent ECUs from communicating, 
a straightforward \emph{Denial of Service (DOS)} attack consists in flooding the bus with frames with the CAN ID set to 0x000. As the CAN ID of a frame determines its priority as mentioned in Section~\ref{sec:can}, repeatedly sending messages with CAN ID 0x000 (i.e.~the highest priority) will prevent the other ECUs from transmitting their frames. Such an attack can put the car in an unstable state~\cite{net_miller2013}.

There are also low level attacks relying on bus signal tempering, as demonstrated in \cite{net_froschle2017}, and researchers have been able to conduct \textit{bus-off} attacks on diverse vehicles \cite{net_cho2016error,net_palanca2017stealth, net_iehira2018}. 
Unlike the attacks discussed previously, they do not require sending entire frames onto the bus.
An attacker can abuse the error handling mechanism of the CAN protocol by sending a few bits at the same time the targeted ECU is transmitting a frame onto the bus.
Consequently the corruption of the message will trigger the fault confinement mechanism of CAN.
After a certain number of errors, the targeted ECU will be put in ``bus off'' mode, preventing it from sending new frames.
The attacker can then freely send her CAN frames, posing as the silenced ECU. 
As these papers explain, these attacks are not detectable by current frame-based NIDS since they do not involve injecting whole CAN frames.

\subsection{Network Intrusion Detection Systems}

In this section, we cover some background information about Network-based Intrusion Detection Systems (NIDS) and their typical attributes used to categorize them, namely their detection method and depth of inspection. Finally we will introduce performance metrics to evaluate these systems.

\subsubsection{Detection method} 
\label{sec:detectionmethod}

There are two main methods used to detect intrusions: \textit{knowledge-based} and \textit{anomaly-based} detection. The later method also encompasses another variant, referred to as \textit{specification-based} detection. In this section we introduce these three methods.

\subsubsection*{Knowledge-based}
\label{sec:knowledge}

Also known as \textit{signature-based} or \textit{misuse detection}, a knowledge-based NIDS uses information about attacks, so-called signatures, as a pattern characterizing a known threat \cite{net_mitchell2014, net_debar2000}. The NIDS will compare the signatures against observed events to identify possible attacks  \cite{net_scarfone2007}. Upon detection (i.e.~an event matches a signature) a action can be executed, typically an alarm will be raised (ideally notifying the network administrator or the security team) and a counter-measure can be initiated (e.g.~termination of the connection) depending on the rule specified for a given attack.

With their black-list approach, knowledge-based NIDS are effective to detect known threats with great accuracy. They generally present a very low rate of false positives since the use of signatures guarantees that each match signifies that a malicious event has been successfully detected. Although, since the set of signatures cannot be exhaustive, such NIDS are unable to catch unknown threats: hackers will find new vulnerabilities to exploit before a signature can be created to detect their attacks (so-called \textit{zero day}). Moreover, some techniques can be employed in order to bypass knowledge-based detection systems \cite{net_colajanni2011}, such as payload encoding.

\subsubsection*{Anomaly-based}
\label{sec:anomaly}

An anomaly-based NIDS first creates a reference model (or ``profile'') of a system by recording and collecting normal/legitimate operations and communications. Then it starts monitoring the current activities of that system. An alert will be generated anytime the NIDS identifies a significant deviation from the model  \cite{net_denning1986}. Building such profiles can be done more or less autonomously using a sample of historical data or even diverse machine learning techniques \cite{net_lunt1992,net_vaccaro1989, net_forrest1996}.

Put simply, anything that does not happen according to the normal behavior previously learned is regarded as an incident. The main advantage of the anomaly-based IDS is twofold: there is no overhead of maintaining a signature database up to date, and they provide the capability to detect unforeseen attacks. Any attack would (in theory) change the normal behavior of a system by, for instance, accessing unusual resources or establishing connections with new machines outside of the trusted network. However this capability comes with the price of false positives: these NIDS are prone to detect legitimate events as malicious if they have not been previously observed during the learning phase \cite{net_scarfone2007}. The quality of the model created will influence the risk of false positives. It is very challenging to build accurate profiles because systems' activities are usually quite complex.

\subsubsection*{Specification-based}
\label{sec:specification}

In the same fashion as anomaly-based detection, specification-based (or ``anomaly-specification-based'') NIDS detect attacks by identifying deviations from a norm \cite{net_sekar2002}. But instead of creating the reference model during an initial learning phase, specifications of a system are manually developed to characterize legitimate program behaviors \cite{net_uppuluri2000}.

Compared to anomaly-based NIDS, the main benefit of this approach is its accuracy in distinguishing legitimate deviations from the malicious ones. If the profile is correctly developed, based on the system's specifications, ``the false positive rate can be comparable to that of knowledge-based detection'' \cite{net_uppuluri2000}. As a matter of fact, since the model is developed manually, it will be more exhaustive than the model created during a definite learning phase. However depending on the system, it can be challenging to retrieve the complete set of specifications.

\subsubsection{Depth of inspection}
\label{sec:depth}

We can distinguished two different levels of depth of inspection. In a \textit{flow-based} approach, a NIDS will look at a collection of packets presenting certain common characteristics. The second approach, referred as \textit{payload-based} (or ``packet-based''), a NIDS will analyze the payload of each packet passing on the network. We will discuss these two methods below.

\subsubsection*{Flow-based inspection}

A flow can be defined as a group of packets sharing common properties and passing an observation point during a certain period \cite{net_rfc7011}. Even if the concept of \textit{flow} is traditionally used in the IT context with TCP/IP communications \cite{net_rfc7011}, Japkowicz and Taylor adapted the term for the CAN bus \cite{net_japkowicz2015}. In their paper, a CAN flow presents a collection of certain characteristics such as the CAN ID and the number of packets contained in that flow.

CAN NIDS researchers have been interested in flow-based detection as this approach fits well CAN traffic. The majority of ECUs in a vehicle communicate at a very specific time. This regularity makes it easy (in theory) to create a model in which communication patterns for each CAN ID are defined. A flow-based NIDS would then look for deviations from these patterns, and would raise an alert when a packet arrives too early or too late compared to the norm.

\subsubsection*{Payload-based inspection}

Payload-based NIDS focus on the payload content of a packet. As discussed in Section \ref{sec:attacks}, certain CAN bus attacks rely on crafting messages with a specific payload. A car hacker can send a packet with a specific CAN ID and a particular payload which can affect the receiving ECUs' behavior. If the attacker manage to send this frame without altering the communication patterns, her attack undetectable by flow-based NIDS.

Detecting these attacks requires to analyze the payload of messages and compare it with either a historical model considered to be legitimate, or with a signature of an attack. In an anomaly-based detection approach, one can create a normal model defining how legitimate payloads should look like, either by learning it or using vendor's specifications, as we discussed in Section \ref{sec:anomaly}. Any deviation in payload values could indicate a potential attack. One could also rely on a knowledge-based method to compare the packet's payload against signatures of known threats, either with string matching algorithms, or by using regular expression matching algorithm \cite{net_elmaghraby2017}. 

%
%
%

\section{Challenges and threat analysis}
\label{sec:threatmod}

In this section, we start by outlining the assumptions scoping our research. For this scope we then provide a threat analysis for the CAN bus in which we detail the capabilities of an adversary and the attacks she can launch.

\subsection{Assumptions}

We assume that the attacker already has a foothold onto the CAN bus of a car. Having access to CAN bus is trivial if the attacker has physical access to the car. She could connect a CAN device to the OBD-II port (present in modern cars to give easy access to mechanics for maintenance operations), and, depending on the architecture of the car~\cite{net_miller2014survey}, she could start transmitting crafted frames. However, car makers tend to disregard such cases and are only concerned with illegal remote access. Researchers have already demonstrated the feasibility of remotely compromising  external interfaces~\cite{net_miller2015remote,net_checkoway2011comprehensive,net_foster2015,net_nie2017}.

\begin{figure}
 \centering
 \includegraphics[width=\columnwidth]{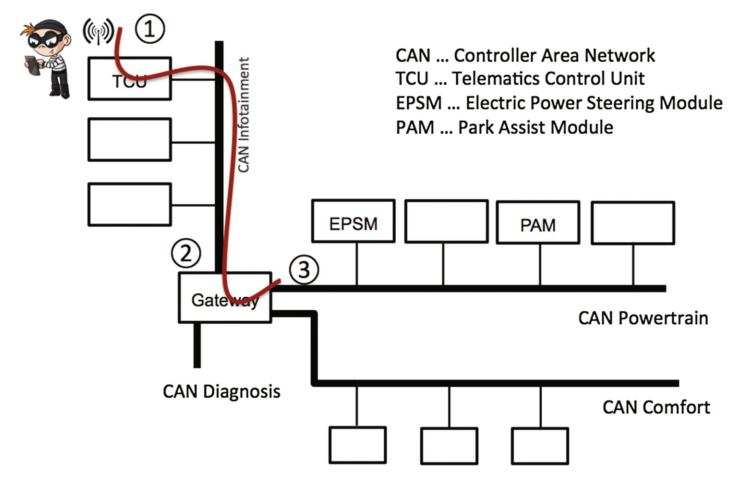}
 \caption{Stages in car hacking, Source: \cite{net_froschle2017}}
 \label{fig:stages}
\end{figure}

There are several ways to obtain remote access to the CAN bus. As explained in~\cite{net_froschle2017}, a possible attack scenario as the one depicted in Fig.~\ref{fig:stages} would involve the following sequence of steps:
\begin{enumerate}[noitemsep]
\item Exploitation of a vulnerability of the telematic unit over the cellular network, granting the attacker remote code execution capabilities and access to the CAN infotainment bus.
\item Pivoting onto the network by compromising the gateway ECU in order to gain access to the safety-critical CAN powertrain bus.
\item Finally the attacker can inject CAN frames onto the bus and perform diverse attacks, as described in Section~\ref{sec:attacks}.
\end{enumerate}
In our technical report we focus on the third step of this scenario. 

\subsection{Threat analysis}
\label{threatanalysis}

In this section we outline the capabilities of an attacker and specify the CAN bus attacks considered in our research.

\subsubsection{Capabilities}
\label{sec:capabilities}

The simplicity of the CAN protocol, as mentioned in Section \ref{sec:can}, offers a number of capabilities to an attacker once she has 
gained access to the network. Because of unencrypted communication, the attacker can read the messages out of the bus and learn about CAN IDs and ECUs' communication patterns in order to reverse engineer the bus. This would allow her to pose as a legitimate ECU by replaying or forging messages with arbitrary CAN ID and payload. Due to the broadcast nature of the protocol, all ECUs on the bus will receive the fake messages, and eventually perform certain actions accordingly. By sending crafted packets set to the specific values, an attacker can for instance steer the car to the direction of her choice, as demonstrated in~\cite{net_miller2013}. The actions she will perform to the car depend on her motivations and intentions.

\subsubsection{Attacks}
\label{sec:threatmod-attacks}

As explained in 
Section~\ref{sec:attacks}, we consider the following attacks:
\begin{enumerate}[noitemsep]
\item Use of diagnostic messages
\item Fuzzing attack
\item Replay/spoofing attack
\item Denial of Service
\end{enumerate}

To address these attacks, researchers have proposed a number of NIDS for CAN networks which aim at detecting injection attacks. In addition, there have been some NIDS proposed to detect 5) \textit{suspension attacks}, in which an ECU would suddenly stop emitting its frames \cite{net_cho2016,net_taylor2016}. While a suspension of frames would actually be more a consequence of another attack (such as a low level attack of \cite{net_palanca2017stealth}), we also include it in our research to see whether frame level NIDS would be able to detect a silenced ECU. In the next section will be presented these NIDS.

\section{Survey of in-vehicle NIDS}
\label{sec:survey}

\subsection{Methodology}

In this section we present the scope of our survey and introduce the different dimensions which will be used in our taxonomy to categorize CAN bus NIDS.

\subsubsection{Paper selection / scope}

For our study we consider all NIDS approaches that perform CAN bus attack detection on a frame level.
%
Some approaches~\cite{net_cho2016,net_murvay2014, net_cho2017, net_choi2018, net_choi2018voltage,net_kneib2018}, while referred to as NIDS, 
actually attempt to authenticate (or fingerprint) ECUs on the bus in order to detect illegitimate sender(s) by using low-level signal characteristics, such as the shape of electrical signals on the bus.
Such approaches use a different attacker model as we consider illegitimate
traffic may come from the ``correct'' but compromised ECU.
Thus, while such approaches are intelligent and seem promising, they would require another dedicated survey and we exclude them from this study.

\subsubsection{Dimensions of the taxonomy}

During our study we realized that the traditional taxonomy for NIDS does not fit very well the particularities of CAN bus NIDS.
NIDS are typically classified depending on the detection method, which can be referred to as \textit{knowledge-based}, using signatures of attacks, or \textit{anomaly-based}, looking for deviation from a system's model. Additionally the later method also encompasses another variant, called \textit{specification-based} detection, where the model of the system is designed based on specifications instead of being learned over time.
While this distinction is generally accurate in IT, CAN bus NIDS call for a different categorization.
To create a well-fitting CAN-specific NIDS categorization, we introduce three dimensions, namely 1) the number of frames, 2) the data used for detection, and 3) how the detection model is built.

\paragraph{Dimension 1 - Number of frames}

This dimension refers to the amount of messages required by a NIDS to detect an attack on the CAN bus. The approaches surveyed propose to detect attacks with a single CAN message, two consecutive messages, or with all the messages contained within a window.

\paragraph{Dimension 2 - Data used}

A CAN bus NIDS will use different features of CAN frames to detect attacks, such as its arbitration ID, and/or its payload. In addition, the timing characteristic of CAN communications can be leveraged to detect attacks by looking at the time interval between frames.

\paragraph{Dimension 3 - Model building}

CAN bus NIDS using an anomaly-based detection approach will likely use a model of the system to be protected and will flag deviations from that model as attacks. There are two main ways to build such models: either by using system's specifications, or by learning it over a certain period of time.

\subsection{Survey}

In this section we give an overview of the papers meeting our study requirements and briefly describe how their proposed system operates.
We structure the following discussion based on the first dimension.

\subsubsection{Single message}
A number of CAN NIDS aim at detecting attacks based on a single frame.
%
NIDSs may learn which frames are legitimate. For example,
\cite{net_kang2016} uses a deep neural network. The model is trained based on the underlying statistical properties of ``normal and hacking CAN packets''. It will then extract the low-dimensional features of CAN frames to discriminate normal packets from malicious ones.
Bloom filtering techniques are used in~\cite{net_groza2019} to assess the periodicity of frames, relying on CAN IDs and certain parts of the payload. This approach allows the detection of replay and modification attacks.
The application of four different fuzzy algorithms on CAN payloads in~\cite{net_martinelli2017} allows to classify CAN frames as being normal or injected.

Instead of learning, NIDSs can rely on the specification. A NIDS may, for example look for CAN IDs that are not part of a specified list of legitimate IDs~\cite{net_ling2012}. %
Abbott-McCune and Shay propose~\cite{net_abbott-mccune2017} detection of illegitimate use of CAN IDs using the fact that only one ECU may send a certain CAN ID. By deploying on the ECUs detectors will know if it was actually the correct
one sending a given CAN frame.
Similarly a detector deployed on a gateway could detect if CAN IDs appear on the wrong subnetwork. A disadvantage of such a host-based approach is the need to change these hosts. Redesigning ECUs is impractical or at least costly for manufacturers.



Some attacks, such as replay, use messages that by themselves may look legitimate. Detecting these requires considering the context of the messages, e.g.~by considering additional messages.

\subsubsection{Two consecutive messages}

Some CAN NIDS leverage the regularity of ECUs communications in order to detect attacks. 
Recall attacks like spoofing 
may require sending frames for a specific CAN ID at a much higher rate than normal for the attack to have the desired effect.
As a result, the time interval between consecutive messages for that CAN ID will shorten.
This has been used in various papers. 


Gmiden et al.~\cite{net_gmiden2016} propose a simple intrusion detection method for CAN network, based on the analysis of messages' time intervals.
Every time a message with a certain CAN ID is sent onto the bus, their algorithm calculates the interval time between this message and the previous message of the same ID. Assuming that each CAN ID has its own regular frequency, if the time interval calculated by the NIDS is less than half of the expected value, an alert is raised.

Moore et al.~\cite{net_moore2017} also exploit the regularity in the timing of CAN communication. For each CAN ID, the NIDS stores the time differences between two successive messages and computes the mean arrival time. In addition, it will register the maximum time difference from the mean, which will be used to define a threshold. An alert will be raised if the time between two packets differs from the expected time by more than the maximum time difference plus 15\% of the mean.

Otsuka and Ishigooka~\cite{net_otsuka2014} observed that the frequency of CAN messages may fluctuate due to collisions with other frames. As a result, NIDS relying on naive frequency analysis can be prone to false positives. In an attempt to improve this method, they propose a ``delayed-decision cycle detection'' method to be deployed on a gateway ECU. The idea can be summarized as follow: each CAN ID has a certain cycle of emission called \textit{T}. If a frame is received by the gateway at a time \textit{a}, the legitimate next frame is expected at the time \textit{a + T}. In the situation where a frame with the same CAN ID would be received before, the ECU will hold the frame and wait until \textit{a + T}. If by this time a new frame with the same ID is received, the NIDS concludes that the previous message received was spoofed. By holding the frame, the risk of misdetection is reduced by not raising an alert at the first sign of deviation.

Marchetti and Stabili~\cite{net_marchetti2017}  
exploit the regularity of CAN bus traffic in a different way by noting that CAN IDs tend to arrive in a certain order.
Their solution learns the recurring patterns of transitions between CAN IDs
within sequences of frames and detects attacks by looking for unlikely transitions.

Attack frames will often have an illegitimate payload.
Stabili et al.~\cite{net_stabili2017} consider the Hamming distance between two consecutive payloads to detect large changes in the payload.
Their NIDS learns the normal rate of change per CAN ID as a range for the Hamming distance between successive packets' payload.
Several attacks, like fuzzing, can cause large changes in this Hamming distance allowing them to be spotted.

Taylor et al.~\cite{net_taylor2016} propose an anomaly detector based on a Long Short-Term Memory (LSTM) recurrent neural network (RNN) to detect attacks. The idea is to train a neural network to predict the next packet's payload. 
Frames are considered malicious if they deviate from the predicated value.

Even if communication is regular, fluctuations in the reception cycle may occur which can lead to false positives~\cite{net_otsuka2014}.
As attacks tend to influence frequencies for some time, 
considering sequences of messages may be less sensitive to such fluctuations.


\subsubsection{Messages in a window}

Sequences of messages, referred to as windows, can be formed by taking a
fixed number of messages with a given CAN ID or all messages within a fixed amount of time (all together or again per CAN ID).
NIDSs can extract features of such windows and use them to detect attacks. 

%
%

Japkowicz et al.~\cite{net_japkowicz2015} propose a frequency-based anomaly 
detector which uses time based windows per CAN ID. 
They use the term \emph{flow} for a windows and its features: the CAN ID, the number of packets, the average Hamming distance and variance of the Hamming distance between successive packet payloads, the average time difference and its variance between successive messages. 
They learn a ``historical model'' using one-second sliding windows and compute \textit{t} test scores to compare new traffic to the model.


Waszecki et al.~\cite{net_waszecki2017automotive} note that CAN messages are not being sent perfectly periodically but are ``periodic with jitter''.
Their NIDS uses arrival curves for each message stream,
capturing the earliest and latest time each message in a window should arrive,
thus accounting for the possible jitter. Messages arriving too late or early (or equivalently too many or too few messages having arrived at some point in time) signals an attack.


Lee et al.~\cite{net_lee2017} describe an active approach where they query some
ECU on the bus. Variation in the response time of the ECU may indicate that there is
an attack on the bus. To detect such variations they consider the ``offset'' (how many messages occur on the bus before the response) and the ``interval'' (response time).
Unusual combinations of offset and interval values are indicators of an attack.


Narayanan et al.~\cite{net_narayanan2016} build a model based on Hidden Markov Models capturing the normal, ``safe'' state of the vehicle.
For detection the NIDS maintains a state which is updated by
activities (certain CAN frames) and looks for activities that
are unlikely given the current and previous state.

Muter and Asaj~\cite{net_muter2011entropy} introduce entropy-based detection to the area of in-vehicle networks. They describe entropy as a ``measure of how much coincidence a given dataset contains''.
They notice that automotive network traffic is 
regular and much more structured than traditional IT traffic: frames are simple, fixed format, uses values with clear bounds and payload that follow some logic depending on the frame's function.
%
%
With this characteristic in mind, change in entropy could signal an attack. The  NIDS from~\cite{net_muter2011entropy} thus 
monitors the entropy in an automotive network.

Marchetti et al.~\cite{net_marchetti2016} notice that the experimental evaluation of the entropy-based NIDS of Mutter et al. is rather limited, as it for instance only considers about 15 seconds of CAN traffic, containing only a single class of CAN messages. In order to evaluate the effectiveness of this approach, they implement it and perform some experiments with various parameters and two types of attacks, namely fuzzing and replay attacks, on the data collected from a 2011 Ford Fiesta.

In a similar approach Wu et al.~\cite{net_wu2018} also develop an entropy-based detection method. It uses a sliding window comprised of a fixed number of messages. Compared to the other entropy-based systems, the authors leverage a Simulated Annealing sliding algorithm in order to find an optimal sliding window parameter. 
Additionally Wang et al.~\cite{net_wang2018} propose an entropy-based NIDS focusing on detection of entropy changes of the CAN ID. While Muter's approach~\cite{net_muter2011entropy} considers the CAN ID as an inseparable vector of 11 bits, this system ``analyzes the entropy change bit by bit''.


The three categories of approaches above are designed to detect different types of attacks. To spot all these attacks one would have to combine different methods.

\subsubsection{Combining approaches}

CAN bus attacks can take different forms. Consequently, ``using only a single algorithm is not enough when considering the need to be able to detect various types of malicious CAN messages'' \cite{net_ujiie2016}. 
%
Miller and Valasek~\cite{net_miller2014survey} implement such a hybrid NIDS. They note that all known CAN injection attacks rely on either CAN diagnostic messages or standard messages sent at higher rates. To detect these attacks they design a NIDS with two detection modules. The first module looks for diagnostic messages while driving. The second one focuses on the frequency of CAN messages. 

A similar NIDS has been proposed by Song et al. in their paper \cite{net_song2016intrusion}. They do not discuss the knowledge part in depth, beside mentioning that diagnostic messages should not being seen on the bus while driving (obvious sign of attack). Regarding the behavior module, they also select the message rate as a significant feature like described in the previous paragraph. The NIDS monitors the time interval of messages, and if an interval between two consecutive packets is shorter than normal, it considers the message to be maliciously injected.

Ujiie et al. \cite{net_ujiie2016} 
propose to combine learning white lists of normal CAN IDs and payload with algorithms to look for cyclic CAN messages sent outside of normal cycles,
non-cyclic messages sent at an abnormal frequency,
or messages otherwise not matching learned historical statistical properties.
%
%

In~\cite{net_studnia2018} the authors derive from models of behavior a set of attacks based on a list of forbidden sequences. Their system then checks whether a frame is compliant with the specifications and is consistent with the current state of the system.

The text above discusses NIDS and the influence of ``number-of-frames''-dimension. Below we also consider the effect of the other dimensions.

\subsection{Discussion}

As seen above the number of frames used has implications on the types of attacks that can be found. The data used also impacts types of detectable attacks.
How models are built influences how easy this process is and how much information it needs. All dimensions also impacts the accuracy of the NIDS.

\subsubsection*{Time interval between messages}
As discussed in the previous section, certain NIDS leverage the regularity of CAN communications in order to determine whether the flow of messages is legitimate or not. Such approaches make sense under the assumption that all CAN frames will be sent at a fixed period.

\subsubsection*{CAN ID}
In the case of attacks relying on using specific CAN ID such as fuzzing attacks, NIDS similar to \cite{net_cho2016} focus on the CAN ID of a frame to assert its maliciousness. The NIDS leveraging the CAN ID of a frame to detect attacks are usually performing analysis of single message, as presented in the previous section.

\subsubsection*{Payload}
Another class of NIDS look at the content of a frame's payload for abnormal values. In traditional IT NIDS, this approach is often referred as \textit{deep packet inspection}, as it analyses the payload instead of solely relying on the header's information to detect attacks. Different CAN bus NIDS have been published, for example \cite{net_stabili2017}.

\subsubsection*{Specified}
Build a reference model can be done by manually specifying the communication pattern, according to the vendors' specifications. The model is guaranteed to be complete and exhaustive enough so that the chances of having false positive is drastically reduced~\cite{net_uppuluri2000}. As first proposed by Larson et al. in~\cite{net_larson2008approach}, this technique could greatly enhance the detection capabilities of NIDS, but it is unfortunately not trivial to obtain such specifications from manufacturers. They are often reluctant to disclose them and keep this information confidential.

\subsubsection*{Learned}
As specifications for cars are difficult to obtain, most CAN NIDS try to built their models during a learning phase. Over a period of time, the NIDS will observe the traffic and will derive communication patterns. The challenge with this approach is to make sure that the learning phase encompasses a broad range of situations to guarantee that the model built is as close as possible to the reality. When failing to do so, it increases the risks of false positives.


\subsubsection*{Further considerations}
While out of scope for this survey, appropriate response to detected 
attacks is also an important issue that needs to be addressed when 
considering practical deployment of CAN NIDS.
It is not trivial to choose a ``safe'' response to a possible attack and
only a few possible options are mentioned in the surveyed papers.
Abbott-McCune and Shay~\cite{net_abbott-mccune2017}  propose blocking of a malicious message 
by sending several dominant bits on the bus which will cause an error and invalidate the message. 
Miller and Valasek~\cite{net_miller2014survey} suggest short circuiting the CAN bus when spotting an attack. This would result in putting the car into ``limp mode'', so the driver could safely stop the car.

%

\begin{table}[th]
   \centering
  	\caption{Inventory of proposed CAN NIDs}
{\tabcolsep=0pt\def\arraystretch{1}
\begin{tabularx}{\columnwidth}{l *{8}{Y}}

\toprule

 & \multicolumn{3}{c}{\textbf{Number of frames}}  
 & \multicolumn{3}{c}{\textbf{Data used}} 
 & \multicolumn{2}{c}{\textbf{Model}} \\
 
\cmidrule(lr){2-4} \cmidrule(l){5-7} \cmidrule(l){8-9}
\rot{\textbf{Papers}}  & \rot{Single frame} & \rot{\begin{tabular}{l}Two consec.\\ frames\end{tabular}}& \rot{\begin{tabular}{l}All frames \\in window\end{tabular}}& \rot{\begin{tabular}{l}Timing of\\frames\end{tabular}}& \rot{CAN ID}& \rot{Payload}& \rot{Specified} & \rot{Learned} \\
\toprule

\cite{net_kang2016} & \faCheck & & & & \faCheck & \faCheck & & \faCheck \\ \hline
\cite{net_groza2019} & \faCheck& & & & \faCheck & \faCheck & & \faCheck \\ \hline
\cite{net_martinelli2017} &\faCheck & & & & & \faCheck& &\faCheck \\ \hline
\cite{net_ling2012} & \faCheck& & & & \faCheck& &\faCheck & \faCircle \\ \hline
\cite{net_abbott-mccune2017} & \faCheck &   &      &       & \faCheck &        & \faCheck & \\ \hline

\cite{net_gmiden2016}       &          & \faCheck & & \faCheck & \faCheck & & & \faCheck \\ \hline
\cite{net_moore2017}        & & \faCheck & & \faCheck & & & & \faCheck \\ \hline
\cite{net_otsuka2014}    & & \faCheck & & \faCheck & & & & \faCheck \\ \hline
\cite{net_marchetti2017} & & \faCheck & & & \faCheck & & & \faCheck  \\ \hline
\cite{net_stabili2017}  &  & \faCheck & &  & & \faCheck & & \faCheck \\ \hline
\cite{net_taylor2016} & & \faCheck & & \faCheck & & \faCheck & & \faCheck \\ \hline

\cite{net_japkowicz2015}    & & & \faCheck & \faCheck & & & & \faCheck \\ \hline
\cite{net_waszecki2017automotive} & & & \faCheck & \faCheck & & & \faCheck & \\ \hline
\cite{net_lee2017} & & & \faCheck & \faCheck & & & \\ \hline
\cite{net_narayanan2016} & & &\faCheck & & &\faCheck & & \faCheck \\ \hline
\cite{net_muter2011entropy} & & & \faCheck & \faCheck & & \faCheck & & \faCheck \\ \hline
\cite{net_marchetti2016} & & & \faCheck & \faCheck & & \faCheck & & \faCheck \\ \hline
\cite{net_wu2018} & & & \faCheck& &\faCheck& & & \faCheck \\ \hline
\cite{net_wang2018} & & & \faCheck& & \faCheck& & & \faCheck\\ \hline

\cite{net_miller2014survey} & \faCheck & \faCheck & & \faCheck & \faCheck & & & \faCheck \\ \hline  
\cite{net_song2016intrusion} & \faCheck & \faCheck & & \faCheck & \faCheck & & - & -  \\ \hline
\cite{net_ujiie2016} & \faCheck & & \faCircle & \faCheck & \faCheck & \faCheck & & \faCheck \\ \hline
\cite{net_studnia2018} & \faCheck & & \faCheck &  &\faCheck &\faCheck &\faCheck &  \\ \hline
\cite{net_daxin2018} & - & - & - & \faCheck & \faCheck & \faCheck & & \faCheck\\

\toprule
\end{tabularx}}
   	\label{table:listpublishedids}
    \caption*{\small \faCheck\ Explicitly mentioned in paper\quad \faCircle\  Implicitly mentioned in paper \quad - Unclear/not mentioned at all} 
\end{table}

\section{Conclusion}
\label{sec:conlusion}

This technical report surveys existing CAN NIDS approaches proposed in the literature.
We categorize them based on the information they use and the way the model is
constructed.
However one does not know how do these systems perform. In future work we will investigate how to compare the NIDS from one another. Doing so will require a unifying testing framework.




\bibliographystyle{IEEEtran}
\bibliography{bibliography}

\begin{thebibliography}{10}
\providecommand{\url}[1]{#1}
\csname url@samestyle\endcsname
\providecommand{\newblock}{\relax}
\providecommand{\bibinfo}[2]{#2}
\providecommand{\BIBentrySTDinterwordspacing}{\spaceskip=0pt\relax}
\providecommand{\BIBentryALTinterwordstretchfactor}{4}
\providecommand{\BIBentryALTinterwordspacing}{\spaceskip=\fontdimen2\font plus
\BIBentryALTinterwordstretchfactor\fontdimen3\font minus
  \fontdimen4\font\relax}
\providecommand{\BIBforeignlanguage}[2]{{%
\expandafter\ifx\csname l@#1\endcsname\relax
\typeout{** WARNING: IEEEtran.bst: No hyphenation pattern has been}%
\typeout{** loaded for the language `#1'. Using the pattern for}%
\typeout{** the default language instead.}%
\else
\language=\csname l@#1\endcsname
\fi
#2}}
\providecommand{\BIBdecl}{\relax}
\BIBdecl

\bibitem{net_checkoway2011comprehensive}
S.~Checkoway, D.~McCoy, B.~Kantor, D.~Anderson, H.~Shacham, S.~Savage,
  K.~Koscher, A.~Czeskis, F.~Roesner, T.~Kohno \emph{et~al.}, ``Comprehensive
  experimental analyses of automotive attack surfaces.'' in \emph{USENIX
  Security Symposium}.\hskip 1em plus 0.5em minus 0.4em\relax San Francisco,
  2011.

\bibitem{net_miller2015remote}
C.~Miller and C.~Valasek, ``Remote exploitation of an unaltered passenger
  vehicle,'' \emph{Black Hat USA}, 2015.

\bibitem{net_miller2016}
C.~Valasek and C.~Miller, ``{CAN Message Injection - OG Dynamite Edition},''
  \url{http://illmatics.com/can message injection.pdf}, 2016.

\bibitem{net_nie2017}
S.~Nie, L.~Liu, and Y.~Du, ``{Free-Fall : Hacking Tesla From Wireless To Can
  Bus},'' \emph{BlackHat USA 2017}, pp. 1--16, 2017.

\bibitem{net_abbott-mccune2017}
S.~Abbott-Mccune and L.~A. Shay, ``{Intrusion prevention system of automotive
  network CAN bus},'' \emph{ICCST 2017}, 2017.

\bibitem{net_gmiden2016}
M.~Gmiden, M.~H. Gmiden, and H.~Trabelsi, ``An intrusion detection method for
  securing in-vehicle can bus,'' in \emph{STA}, 2016, pp. 176--180.

\bibitem{net_moore2017}
M.~R. Moore, R.~A. Bridges, F.~L. Combs, M.~S. Starr, and S.~J. Prowell,
  ``Modeling inter-signal arrival times for accurate detection of can bus
  signal injection attacks,'' in \emph{CISRC}, 2017.

\bibitem{net_japkowicz2015}
N.~Japkowicz, A.~Taylor, and S.~Leblanc, ``Frequency-based anomaly detection
  for the automotive can bus,'' in \emph{World Congress on Industrial Control
  Systems Security (WCICSS)}, 2015, pp. 45--49.

\bibitem{net_otsuka2014}
S.~Otsuka and T.~Ishigooka, ``{CAN Security : Cost-Effective Intrusion
  Detection for Real-Time Control Systems Overview of In-Vehicle Networks},''
  \emph{SAE Technical Paper}, 2014.

\bibitem{net_waszecki2017automotive}
P.~Waszecki, P.~Mundhenk, S.~Steinhorst, M.~Lukasiewycz, R.~Karri, and
  S.~Chakraborty, ``{Automotive E/E Architecture Security via Distributed
  In-Vehicle Traffic Monitoring},'' \emph{IEEE Transactions on Computer-Aided
  Design of Integrated Circuits and Systems}, 2017.

\bibitem{net_cho2016}
K.-t. Cho and K.~G. Shin, ``{Fingerprinting Electronic Control Units for
  Vehicle Intrusion Detection},'' \emph{Proc. of the 25th USENIX Security
  Symposium}, pp. 911--927, 2016.

\bibitem{net_muter2011entropy}
M.~M{\"u}ter and N.~Asaj, ``Entropy-based anomaly detection for in-vehicle
  networks,'' in \emph{IV 2011, IEEE}, 2011, pp. 1110--1115.

\bibitem{net_marchetti2016}
M.~Marchetti, D.~Stabili, A.~Guido, and M.~Colajanni, ``{Evaluation of anomaly
  detection for in-vehicle networks through information-theoretic
  algorithms},'' \emph{IEEE 2nd International Forum on Research and
  Technologies for Society and Industry Leveraging a Better Tomorrow}, 2016.

\bibitem{net_stabili2017}
D.~Stabili, M.~Marchetti, and M.~Colajanni, ``{Detecting attacks to internal
  vehicle networks through Hamming distance},'' \emph{AEIT 2017, IEEE}, 2017.

\bibitem{net_taylor2016}
A.~Taylor, S.~Leblanc, and N.~Japkowicz, ``{Anomaly Detection in Automobile
  Control Network Data with Long Short-Term Memory Networks},'' \emph{DSAA
  2016}, pp. 130--139, 2016.

\bibitem{net_miller2014survey}
C.~Miller and C.~Valasek, ``A survey of remote automotive attack surfaces,''
  \emph{Black Hat USA}, 2014.

\bibitem{net_song2016intrusion}
H.~M. Song, H.~R. Kim, and H.~K. Kim, ``Intrusion detection system based on the
  analysis of time intervals of can messages for in-vehicle network,'' in
  \emph{ICOIN 2016}.\hskip 1em plus 0.5em minus 0.4em\relax IEEE, 2016, pp.
  63--68.

\bibitem{net_ujiie2016}
Y.~Ujiie, T.~Kishikawa, T.~Haga, H.~Matsushima, T.~Wakabayashi, M.~Tanabe,
  Y.~Kitamura, and J.~Anzai, ``{A Method for Disabling Malicious CAN Messages
  by Using a CMI-ECU},'' 2016.

\bibitem{net_lee2017}
H.~Lee, S.~H. Jeong, and H.~K. Kim, ``{OTIDS : A Novel Intrusion Detection
  System for In-vehicle Network by using Remote Frame},'' \emph{PST 2017},
  2017.

\bibitem{net_marchetti2017}
M.~Marchetti and D.~Stabili, ``{Anomaly detection of CAN bus messages through
  analysis of ID sequences},'' \emph{IV 2017, IEEE}, pp. 1577--1583, 2017.

\bibitem{net_kang2016}
M.~J. Kang and J.~W. Kang, ``{A novel intrusion detection method using deep
  neural network for in-vehicle network security},'' \emph{VTC 2016}, 2016.

\bibitem{net_larson2008approach}
U.~E. Larson, D.~K. Nilsson, and E.~Jonsson, ``An approach to
  specification-based attack detection for in-vehicle networks,'' in
  \emph{Intelligent Vehicles Symposium, 2008 IEEE}.\hskip 1em plus 0.5em minus
  0.4em\relax IEEE, 2008, pp. 220--225.

\bibitem{net_hoppe2011}
T.~Hoppe, S.~Kiltz, and J.~Dittmann, ``{Security threats to automotive CAN
  networks -- Practical examples and selected short-term countermeasures},''
  \emph{Reliability Engineering and System Safety}, vol.~96, no.~1, pp. 11--25,
  2011.

\bibitem{net_muter2010structured}
M.~M{\"u}ter, A.~Groll, and F.~C. Freiling, ``A structured approach to anomaly
  detection for in-vehicle networks,'' in \emph{IAS 2010}.\hskip 1em plus 0.5em
  minus 0.4em\relax IEEE, 2010, pp. 92--98.

\bibitem{net_muter2011}
M.~M{\"{u}}ter, A.~Groll, and F.~C. Freiling, ``{Anomaly Detection for
  In-Vehicle Networks using a Sensor-based Approach},'' \emph{Journal of
  Information Assurance and Security}, vol.~6, pp. 132--140, 2011.

\bibitem{net_introtocan}
S.~Corrigan, ``{Introduction to the Controller Area Network (CAN)},'' Internet
  Requests for Comments, {Texas Instruments}, Tech. Rep., August 2002.

\bibitem{net_understanding_can}
M.~D. Natale, \emph{Understanding and using the Controller Area Network}, 2008.

\bibitem{net_canproto_understandingcan}
``{CAN Protocol - Understanding the Controller Area Network Protocol},''
  \url{https://www.engineersgarage.com/article/what-is-controller-area-network},
  accessed: 2018-07-25.

\bibitem{net_can_spec}
BOSCH, \emph{CAN Specification Version 2.0}, 1991.

\bibitem{net_canoverview}
``{CAN Overview},'' \url{http://www.ni.com/white-paper/2732/en/}, accessed:
  2018-07-25.

\bibitem{net_cantuto}
``{CAN tutorial},'' {Contemporary Controls}, Tech. Rep.

\bibitem{net_rouf2010}
I.~Rouf, R.~Miller, H.~Mustafa, T.~Taylor, S.~Oh, W.~Xu, M.~Grutese, W.~Trappe,
  and I.~Seskar, ``{Security and privacy vulnerabilities of in-car wireless
  networks: A tire pressure monitoring system case study.}'' \emph{Proc. of the
  USENIX Security Symposium}, vol.~39, no.~4, pp. 11--13, 2010.

\bibitem{net_miller2013}
\BIBentryALTinterwordspacing
C.~Miller and C.~Valasek, ``{Adventures in Automotive Networks and Control
  Units},'' \emph{Hacktivity 2015}, 2013. [Online]. Available:
  \url{http://illmatics.com/car{\_}hacking.pdf}
\BIBentrySTDinterwordspacing

\bibitem{net_froschle2017}
S.~Froschle and A.~Stuhring, ``{Analyzing the Capabilities of the CAN
  Attacker},'' \emph{ESORICS 2017}, vol. 10492, pp. 464--482, 2017.

\bibitem{net_cho2016error}
K.-t. Cho and K.~G. Shin, ``{Error Handling of In-vehicle Networks Makes Them
  Vulnerable},'' \emph{CCS}, pp. 1044--1055, 2016.

\bibitem{net_palanca2017stealth}
A.~Palanca, E.~Evenchick, F.~Maggi, and S.~Zanero, ``A stealth, selective,
  link-layer denial-of-service attack against automotive networks,'' in
  \emph{International Conference on Detection of Intrusions and Malware, and
  Vulnerability Assessment}.\hskip 1em plus 0.5em minus 0.4em\relax Springer,
  2017, pp. 185--206.

\bibitem{net_iehira2018}
K.~{Iehira}, H.~{Inoue}, and K.~{Ishida}, ``Spoofing attack using bus-off
  attacks against a specific ecu of the can bus,'' in \emph{15th IEEE Annual
  Consumer Communications Networking Conference (CCNC)}, 2018, pp. 1--4.

\bibitem{net_mitchell2014}
R.~Mitchell, I.-r. Chen, and V.~Tech, ``{A Survey of Intrusion Detection
  Techniques for Cyber-Physical Systems},'' vol.~46, no.~4, 2014.

\bibitem{net_debar2000}
H.~Debar, M.~Dacier, and A.~Wespi, ``{A revised taxonomy for
  intrusion-detection systems},'' pp. 361--378, 2000.

\bibitem{net_scarfone2007}
K.~Scarfone and P.~Mell, ``{Guide to Intrusion Detection and Prevention Systems
  (IDPS) Recommendations of the National Institute of Standards and
  Technology},'' \emph{Nist Special Publication}, vol. 800-94, p. 127, 2007.

\bibitem{net_colajanni2011}
M.~Colajanni, L.~{Dal Zotto}, M.~Marchetti, and M.~Messori, ``{The problem of
  NIDS evasion in mobile networks},'' \emph{2011 4th IFIP International
  Conference on New Technologies, Mobility and Security, NTMS 2011 - Proc.},
  2011.

\bibitem{net_denning1986}
E.~Denning, R.~Ave, and M.~Park, ``{An intrusion detection model},'' \emph{IEEE
  Transactions on Software Engineering}, pp. 118--131, 1986.

\bibitem{net_lunt1992}
T.~F. Lunt, A.~Tamaru, F.~Gilham, N.~R. Jagan, C.~Jalali, and P.~G. Neumann,
  ``{A real-time intrusion-detection expert system (ides)},'' 1992.

\bibitem{net_vaccaro1989}
H.~Vaccaro and G.~Liepins, ``{Detection of anomalous computer session
  activity},'' 1989.

\bibitem{net_forrest1996}
S.~Forrest, S.~Hofmeyr, A.~Somayaji, and T.~Longstaff, ``{A sense of self for
  Unix processes},'' \emph{Proc. 1996 IEEE Symposium on Security and Privacy},
  pp. 120--128, 1996.

\bibitem{net_sekar2002}
R.~Sekar, A.~Gupta, J.~Frullo, T.~Shanbhag, A.~Tiwari, H.~Yang, and S.~Zhou,
  ``{Specification-based anomaly detection},'' \emph{Proc. of the 9th ACM
  conference on Computer and communications security - CCS '02}, vol.~26,
  no.~2, p. 265, 2002.

\bibitem{net_uppuluri2000}
P.~Uppuluri and R.~Sekar, ``{Experiences with specification-based intrusion
  detection},'' \emph{RAID 2000}, pp. 1--18, 2000.

\bibitem{net_rfc7011}
B.~Claise, B.~Trammell, and P.~Aitken, ``{Specification of the IP Flow
  Information Export (IPFIX) Protocol for the Exchange of Flow Information},''
  Internet Requests for Comments, {Internet Engineering Task Force (IETF)},
  {RFC} 7011, September 2013.

\bibitem{net_elmaghraby2017}
R.~T. El-Maghraby, N.~M.~A. Elazim, and A.~M. Bahaa-Eldin, ``A survey on deep
  packet inspection,'' in \emph{2017 12th International Conference on Computer
  Engineering and Systems (ICCES)}, Dec 2017, pp. 188--197.

\bibitem{net_foster2015}
I.~Foster, A.~Prudhomme, K.~Koscher, and S.~Savage, ``Fast and vulnerable: A
  story of telematic failures,'' in \emph{9th {USENIX} Workshop on Offensive
  Technologies ({WOOT} 15)}.\hskip 1em plus 0.5em minus 0.4em\relax Washington,
  D.C.: {USENIX} Association, 2015.

\bibitem{net_murvay2014}
P.-S. Murvay and B.~Groza, ``{Source Identification Using Signal
  Characteristics in Controller Area Networks},'' \emph{IEEE Signal Processing
  Letters}, vol.~21, no.~4, 2014.

\bibitem{net_cho2017}
K.-T. Cho and K.~Shin, ``{Viden: Attacker Identification on In-Vehicle
  Networks},'' \emph{CoRR}, 2017.

\bibitem{net_choi2018}
W.~Choi, H.~J. Jo, S.~Woo, J.~Y. Chun, J.~Park, and D.~H. Lee, ``{Identifying
  ECUs Using Inimitable Characteristics of Signals in Controller Area
  Networks},'' \emph{IEEE Transactions on Vehicular Technology}, vol.~67,
  no.~6, pp. 4757--4770, 2018.

\bibitem{net_choi2018voltage}
W.~Choi, K.~Joo, H.~J. Jo, M.~C. Park, and D.~H. Lee, ``{VoltageIDS : Low-Level
  Communication Characteristics for Automotive Intrusion Detection System},''
  \emph{IEEE Transactions on Information Forensics and Security}, vol.~13,
  no.~8, pp. 2114--2129, 2018.

\bibitem{net_kneib2018}
M.~Kneib and C.~Huth, ``Scission: Signal characteristic-based sender
  identification and intrusion detection in automotive networks,'' in
  \emph{Proceedings of the 2018 ACM SIGSAC Conference on Computer and
  Communications Security}, 2018, pp. 787--800.

\bibitem{net_groza2019}
B.~Groza and P.~S. Murvay, ``{Efficient Intrusion Detection with Bloom
  Filtering in Controller Area Networks},'' \emph{IEEE Transactions on
  Information Forensics and Security}, vol.~14, no.~4, pp. 1037--1051, 2019.

\bibitem{net_martinelli2017}
F.~Martinelli, F.~Mercaldo, V.~Nardone, and A.~Santone, ``{Car hacking
  identification through fuzzy logic algorithms},'' \emph{FUZZ-IEEE 2017},
  2017.

\bibitem{net_ling2012}
C.~Ling and D.~Feng, ``{An Algorithm for Detection of Malicious Messages on CAN
  Buses},'' 2012.

\bibitem{net_narayanan2016}
S.~N. Narayanan, S.~Mittal, and A.~Joshi, ``{OBD SecureAlert : An Anom\-aly
  Detection System for Vehicles},'' \emph{SMARTCOMP 2016, IEEE}, 2016.

\bibitem{net_wu2018}
W.~Wu, Y.~Huang, R.~Kurachi, G.~Zeng, G.~Xie, R.~Li, and K.~Li, ``{Sliding
  Window Optimized Information Entropy Analysis Method for Intrusion Detection
  on In-Vehicle Networks},'' \emph{IEEE Access}, vol.~6, pp. 45\,233--45\,245,
  2018.

\bibitem{net_wang2018}
Q.~Wang, Z.~Lu, and G.~Qu, ``{An Entropy Analysis based Intrusion Detection
  System for Controller Area Network in Vehicles},'' \emph{CoRR}, 2018.

\bibitem{net_studnia2018}
I.~Studnia, Y.~Laarouchi, M.~Ka{\^{a}}niche, V.~Nicomette, and E.~Alata, ``{A
  language-based intrusion detection approach for automotive embedded
  networks},'' \emph{International Journal of Embedded Systems}, vol.~10, 2018.

\bibitem{net_daxin2018}
D.~Tian, Y.~Li, Y.~Wang, X.~Duan, C.~Wang, W.~Wang, R.~Hui, and P.~Guo, ``An
  intrusion detection system based on machine learning for {CAN}-bus,'' in
  \emph{INISCOM 2018}.\hskip 1em plus 0.5em minus 0.4em\relax Springer, 2018,
  pp. 285--294.

\end{thebibliography}

\end{document}